# Fundamental Principle of Information-to-Energy Conversion

ANTONIO ALFONSO-FAUS
Department of Aerotecnia
Escuela de Ingeniería Aeronáutica y del Espacio, Madrid Technical University
Plaza del Cardenal Cisneros, 3, 28040 Madrid
SPAIN
E-mail address: aalfonsofaus@yahoo.es

*Abstract.* - The equivalence of 1 bit of information to entropy was given by Landauer in 1961 as kln2, k the Boltzmann constant. Erasing information implies heat dissipation and the energy of 1 bit would then be (the Landauer´s limit) kT ln 2, T being the ambient temperature. From a quantum-cosmological point of view the minimum quantum of energy in the universe corresponds today to a temperature of ~$10^{-29}$ °K, probably forming a cosmic background of a Bose condensate [1]. Then, the bit with minimum energy today in the Universe is a quantum of energy ~ $10^{-45}$ ergs, with an equivalent mass of ~ $10^{-66}$ g. Low temperature implies low energy per bit and, of course, this is the way for faster and less energy dissipating computing devices. Our conjecture is this: the possibility of a future access to the CBBC (a coupling/channeling?) would mean a huge jump in the performance of these devices.

## 1 Introduction

Logical operations in computing imply erasing bits. This also implies energy dissipation. In 1961 Landauer [2] established the equivalence of one bit of information with entropy as k ln 2, k the Boltzmann constant. At room temperature T, this is equivalent to an energy kT ln 2, a bound known as the Landauer´s limit. Later [3] Landauer stated *"Information is physical"*. The question is: can we lower the amount of energy dissipated by approaching to the ideal limit of reversible computation (no bit erasing)?

Lloyd [4] stated that *"Merely by existing all physical systems register information"*. More recently Bérut [5] experienced the existence of the Landauer´s bound, demonstrating the *"intimate link between information theory and thermodynamics"*. It appears that the Landauer´s statement that *"information is physical"* has been successfully experimented. One has to remember J.A. Wheeler famous statement *"it from bit"*, meaning (in our interpretation) that bits are the underpinning of the universe.

Going back to 1929, Szilárd [6] established a protocol for a feedback control mechanism in order to approach to the theoretical limit of reversible computation, no energy dissipation (Szilárd demon). But we know today that efforts to achieve this limit imply worse and worse computing performance. So, some erasing of bits is always necessarily present. In any isolated system, with no heat going in or out, an irreversible process inside it implies that the entropy necessarily increases. On the other hand, conversion of information to free energy has been theoretically clarified as possible [7] and experimental demonstrations of information-to-energy conversion do exist [7], [8]. Then, a big reservoir of information inside a system can be the source of energy to be transferred to any part of this system. If there is a cosmic background Bose condensate of information bits in the universe[1] it may be its source of energy.

We will determine the quantum of minimum energy (a self gravitational potential energy $E_g$) related to any quantum particle, photons, baryons or dark components of the universe. We have proposed a cosmic background Bose condensate (CBBC), presented elsewhere [1], and in this work we identify the information bits with these quanta. We then conjecture the possibility of faster and lower dissipating computing devices if channeling and/or coupling with the CBBC becomes possible in the future. It may be that the universe works in this way, being underpinned by fast and exact computation.

## 2 The quantum of minimum energy

One may ask if there is a quantum of minimum energy in the universe. Theoretically the answer is yes, the quantum of self gravitational potential energy physically tied to any quantum particle on can think of: photons, baryons, possible dark matter and/or dark energy quanta, or any other field in the universe.

In 1972 Weinberg [9] advanced a clue to suggest that large numbers are determined by both, microphysics and the influence of the whole universe. This is a hope to connect in some way general relativity and quantum mechanics. He constructed a mass using the physical constants G, $\hbar$, c and the Hubble parameter H. This mass was not too different from the mass of a typical elementary particle (like a pion) and is given by

$$m \approx (\hbar^2 H/Gc)^{1/3} \quad (1)$$

In our work here we consider a general elementary particle of mass $m$, the fundamental particle of any field we may consider. These general fields must have gravitational properties, pulling or pushing. So, this particle may include not only baryons and photons but the possible quantum masses of dark matter and dark energy in the universe. Since the mass $m$ disappears from the resultant relation that we are going to find, the conclusion is totally independent on the kind of elementary particle that we may consider, and therefore from the kind of field considered.

We have presented [10] elsewhere an equivalent formula like (1) but instead of using H/c in (1) we use 1/R, R the cosmological scale factor that is of the order of c/H:

$$m \approx (\hbar^2/GR)^{1/3} \quad (2)$$

The self gravitational potential energy $E_g$ of this quantum of mass $m$ (and size its Compton wavelength $\hbar/mc$) is given by

$$E_g = Gm^2 / (\hbar/mc) = Gm^3c/\hbar \quad (3)$$

Combining (1) and (2) we eliminate the mass $m$ to obtain

$$E_g \approx H\hbar = \hbar c/R \quad (4)$$

In this final formulation, $E_g$ being a quantum of gravitational potential energy, there is no G, the gravitational constant. Only relativity, $c$ the speed of light and quantum mechanics: $\hbar$ is Planck's constant usually interpreted as the smallest quantum of action (angular momentum). Since H is of the order of 1/t, t the age of the universe (t being a maximum time today), (4) is the lowest quantum energy state that it may exist. It is equivalent to $\hbar c/\lambda$ with $\lambda \approx$ R of the order of the size of the visible universe (it is the lowest quantum energy state with $\lambda \approx$ ct). It is unlocalized in the visible universe, like the gravitational field. We identify it with the quantum of the self gravitational potential energy of any quantum particle [1], a minimum quantum of energy [15]. We also identify it with the bit, the unit of information [10]. Remembering the statements of Lloyd [4] and Landauer [5]: "*Merely by existing, all physical systems register information*" and "*Information is physical*", today we say here [1]:

<u>All physical systems of mass M (energy Mc$^2$) are equivalent to an amount of information in number of bits of the order of</u>

*Number of bits* $\approx Mc^2/E_g \approx Mc^2/(H\hbar) \approx McR/\hbar$ (5)

This is our proposed principle of information-to-energy conversion: the bit of information has an energy $E_g$ given in (4). For the visible universe this is about $10^{121}$ bits. For galaxies is about $10^{110}$ bits. For the sun is about $10^{99}$ bits. For the earth is about $10^{93}$ bits, and for the human brain about $10^{70}$ bits, and so on ($10^{43}$ bits for the proton).

## 3 The Cosmic background Bose condensate (CBBC): underpinning of the universe

The energy of the bit of information in (4) gives a minimum energy boson, and a minimum equivalent mass because it has the longest wavelength possible, the maximum size of the visible universe c/H.

The physical property that defines a Bose condensate is "degeneracy" in the energy per particle formulation [1]:

$E/N = 3/2\ kT\ \{\ 1 - 1/2^{5/2}\ Nh^3/V(2\pi mkT)^{3/2}\}$ (6)

The degeneracy is given by the expression after the number 1 in (6). The way to approach a Bose condensate state is then to have a big number of quanta in the same lowest energy state, with very low equivalent mass m and a very low energy kT. An estimate of the degeneracy factor applied to the universe in (6), for our case of a bit of information, gives

$$1/2^{5/2} Nh^3 /V(2\pi mkT)^{3/2} \sim ( h /mc)^3/R^3 \sim O(1) \qquad (7)$$

Then the total energy E is of the order of $NkT$ with N very large ($\sim 10^{121}$) and T very small ($\sim 10^{-29}$ °K).

## 4 Number of bits in the universe

The computational capacity of the universe has been presented by Lloyd [4]. In 1981 Bekenstein [11] found an upper bound for the ratio of the entropy $S_B$ to the energy $E = Mc^2$ of any bounded system with effective size R:

$$S_B /E < 2\pi k R/\hbar c \qquad (8)$$

About ten years later [12], [13], a holographic principle was proposed giving a bound for the entropy $S_h$ of a bounded system of effective size R as

$$S_h \leq \pi k c^3 R^2/\hbar G \qquad (9)$$

We also have the expression for the entropy of a black hole given by the Hawking relation [14]

$$S_H = 4\pi k/\hbar c \ GM^2 \qquad (10)$$

By equating the two bounds in (8) and (9) one gets the Hawking entropy (10), and substituting the values for the universe $R \approx 10^{28}$cm and $M \approx 10^{56}$g we get for the universe today the entropy $S_u$,

$$S_u \approx 10^{121} k \qquad (11)$$

and this is the number of information bits as obtained from (5) for the universe too. Hence, the entropy of a system in units of k, the Boltzmann constant, is the same as the number of bits it has and its total energy is also given by this number in units of $E_g$.

## 5 Demons: Slizárd, Maxwell and our universal expansion

In the introduction we have mentioned that Szilárd [6] established a protocol for a feedback control mechanism (Slizárd demon). This is similar to the later Maxwell´s demon increasing energy from one system to another, with an apparent violation of the second law of thermodynamics. In the case of Slizárd he wanted to approach to the theoretical limit of reversible computation, no energy dissipation. In our case, applying what we have said to the case of the universe, we have an isolated (no heat in or out) expanding universe irreversibly. Then the entropy increases, and therefore the number of bits. From the holographic principle we know that this numbers go as $R^2$ which in units of Planck´s length is about $10^{121}$. This is equivalent to an energy of about $10^{121} E_g$, which is about the present total energy of the universe, $Mc^2$. The question is, where does it come from? It increases as $R^2$ and R is increasing because the universe is expanding.

To the well known entropy problem, the origin of such big entropy, we identify here that we also have an energy problem. And therefore an information problem too. We know that an isolated adiabatic system, evolving in an irreversible way, necessarily has its entropy increasing. And we see here that its information content is also increasing. For the universe we advance a conjecture: one way to solve this conundrum is to postulate the existence of a big reservoir of Bose condensate as defined elsewhere [1], and here in section 3.

## 6 Conclusions

In this work we present the following items that may have a deep significance:

The bit of information is equivalent to a quantum of minimum energy that, at the room temperature T, is given by $E_g \approx kT \approx \hbar c/R$. It is defined in terms of the size of the universe R. Therefore the temperature T follows the well known cosmological relation T ÷ 1/R.

We conjecture the existence of a big reservoir, of constant energy density, as a cosmic background of a Bose condensate (CBBC). This is a background of pure information, a calm superfluid with no viscosity and composed of a high number of quanta all in a minimum energy state. This would explain for the universe the expansion, as a Maxwell-type demon converting information into energy. Such a reservoir has an extremely low temperature, today about $10^{-29}$ °K, and we conjecture that any future possibility of interaction with our room temperature ambient would imply a huge jump in the performance of computing devices.

*References:*
[1] Alfonso-Faus, A., Fullana i Alfonso, M.J., 2013, *"Cosmic Background Bose Condensation (CBBC)"* Astrophysics and Space Science, DOI **10.1007/s10509-013-1500-8** May, 2013. And arXiv:1306.3468
[2] Landauer, R., *"Irreversibility and heat generation in the computing process"*, IBM J. Res. Dev. 5, 183-191 (1961).
[3] Landauer, R., *"Dissipation and noise immunity in computation and communication"*, NATURE VOL. 335, 779-784, 27 October 1988.
[4] Lloyd, S., *"Ultimate physical limits to computation"*. Nature **406,** 1047-1054 (2000), and in *"Computational capacity of the universe"*, arXiv quant-ph/0110141v1, 24 Oct 2001.
[5] Bérut, A. et al., *"Experimental verification of Landauer´s principle linking information and thermodynamics"*, Nature **483,** 187-189 (08 March 2012)
[6] Zilárd, L., *"On the Decrease in Entropy in a Thermodynamic System by the Intervention of Intelligent Beings"* (Uber die Entropieverminderung in einem thermodynamischen System bei Eingriffen intelligenter Wesen), *Zeitschrift fur Physik***, 53**, 840-56. (1929).
[7] Toyabe, S. et al., *"Experimental demonstration of information-to-energy conversion and validation of the generalized Jarzynski equality"*, Letters NATURE PHYSICS I VOL 6 December 2010
[8] Funo., K., et al., *"Thermodynamic Work Gain from Entanglement"*, arXiv:1207.6872v2 [quant-ph] 14 Nov 2012
[9**]** S. Weinberg, *Gravitation and Cosmology: Principles and Applications of the General Theory of Relativity*, John Wiley & Sons Inc. (1972)
[10] Alfonso-Faus, A., (2011) *"Quantum gravity and information theories linked by the physical properties of the bit"*, arXiv: 1105.3143
[11] Bekenstein, J.D., (1981), *"Universal upper bound on the entropy-to-energy ratio for bounded systems"*, *Phys. Rev. D* **23**, number 2, p. 287-298.
[12] 't Hooft, G., (1993). *"Dimensional Reduction in Quantum Gravity"*. *Preprint*. arXiv:gr-qc/9310026.
[13] Susskind, L., (1995) *"The World as a hologram,"* *J. Math. Phys*. **36**, 6377 and arXiv:hep-th/9409089.
[14] Hawking, S.W., (1974), *"Black hole explosions?" Nature* 248, 30
[15] Alfonso-Faus, A., *"Universality of the self gravitational potential energy of any fundamental particle"*, arXiv:1107.3426 and "Astrophysics and Space Science", 337:363-365 (2012).